\documentclass[12pt,preprint]{aastex}

\usepackage{graphicx}

\shorttitle{Giant Planets and Parent Bodies of Iron Meteorites}
\shortauthors{Haghighipour \& Scott}

\begin{document}

\title{On The Effect of Giant Planets on the Scattering of Parent Bodies \\
of Iron Meteorite from the Terrestrial Planet Region into the \\
Asteroid Belt: A Concept Study}

\author{Nader Haghighipour}
\affil{Institute for Astronomy and NASA Astrobiology Institute,
University of Hawaii-Manoa, Honolulu, HI 96822, USA}

\email{nader@ifa.hawaii.edu}

\and 

\author{Edward R. D. Scott}
\affil{Hawaii Institute for Geophysics and Planetology, University
of Hawaii-Manoa, Honolulu, HI 96822}

\begin{abstract}

In their model for the origin of the parent bodies of
iron meteorites, Bottke et al proposed differentiated planetesimals that were 
formed in the region of 1-2 AU during the first 1.5 Myr, as the parent bodies, and
suggested that these objects and their fragments were scattered 
into the asteroid belt as a result of interactions with planetary embryos. Although viable,
this model does not include the effect of a giant planet that might have existed or been growing
in the outer regions. 
We present the results of a concept study where we have examined the effect of a
planetary body in the orbit of Jupiter on the early scattering of planetesimals from 
terrestrial region into the asteroid belt.
We integrated the orbits of a large battery of planetesimals in a disk of planetary embryos,
and studied their evolutions for different values of the mass of the planet.
Results indicate that when the mass of the planet is smaller than 10 Earth-masses, its effects 
on the interactions among planetesimals and planetary embryos is negligible.
However, when the planet mass is between 10 and 50 Earth-masses, simulations point 
to a transitional regime with $\sim$ 50 Earth-mass being the value for which the perturbing 
effect of the planet can no longer be ignored. Simulations also show that further 
increase of the mass of the planet strongly reduces the efficiency of the scattering 
of planetesimals from the terrestrial planet region into the asteroid belt. 
We present the results of our simulations and discuss their possible implications 
for the time of giant planet formation.
\end{abstract}


\section{Introduction}

It has been suggested by \citet{Bottke06} that the extraordinary number of different parent
bodies of iron meteorites, and the dearth of asteroids and meteorites derived from the silicate
mantles and crusts of these objects, could be better understood if these bodies accreted 
not in the asteroid belt, but in the terrestrial planet region (1-2 AU).
If the parent bodies of iron meteorites had formed and differentiated igneously in the main 
asteroid belt, we should have been able to find many metallic asteroids composed of debris from 
cores, as well as numerous olivine-rich asteroids and meteorites composed of mantle 
material. \citet{Burbine96} attributed the lack of olivine-rich and metal-rich 
asteroids to numerous impacts that pulverized the differentiated asteroids.
However, this explanation appears inconsistent with the survival of Vesta's basaltic crust, and
the existence of some unrelated V-type asteroids \citep{Moskovitz08} and
basaltic meteorites that are not from Vesta. 

A major argument of \citet{Bottke06} for suggesting that the igneously differentiated asteroids 
formed at 1-2 AU is that parent bodies of iron meteorites accreted in the first 1.5 Myr
\citep{Kleine05,Markowski06a,Markowski06b} when $^{26}$Al, which has a half-life of 0.73 Myr,
was still capable of melting bodies larger than 20 km in radius \citep{Hevey06}. Accretion
rates would have been much faster at 1-2 AU than in the asteroid belt due to the higher density
of solids and the shorter orbital periods. Later accretion of planetesimals in the asteroid belt
at 2-4 Myr, when $^{26}$Al had largely decayed, accounts for the preponderance of unmelted asteroids
in the belt and the relatively young ages of chondrules \citep[e.g.,][]{Scott08}.

\citet{Bottke06} proposed that igneously differentiated asteroids and their surviving fragments
were dynamically excited by planetary embryos at 1-2 AU and scattered into the asteroid belt 
where they were captured by planetary embryos in that region. To test this hypothesis with
numerical simulations, these authors made many simplifying assumptions to track thousands of massless
test bodies that interacted with a swarm of Moon- to Mars-sized planetary embryos for 10 Myr.
They found that 10\% of the test bodies originally at 1.5-2.0 AU were scattered into the main
belt after 1 Myr. For the 1.0-1.5 AU zone, 1-2\% were injected into the main belt after 2 Myr,
while for the 0.5-1.0 AU zone, 0.01-0.1\% reached the main belt after 6 Myr. More realistic
simulations of Bottke et al. model have not been published.

When the model of \citet{Bottke06} for differentiated planetesimals was proposed, there was
no meteorite evidence suggesting that differentiated asteroids had been disrupted within a few
million years after they formed (although disruption at this time prior to emplacement in the
asteroid belt was a key part of their model).
However, evidence supporting the early disruption of iron meteorite parent bodies has since been 
obtained from cooling rates that were determined from Fe-Ni phases and U-Pb age 
dating. The cooling rates of group IVA irons are correlated with their bulk Ni 
and vary by a factor of $\sim$50. This implies that these objects solidified and cooled 
with little or no silicate mantle \citep{Yang10a} possibly after a grazing 
collision with larger bodies that separated their core and mantle materials
\citep{Asphaug06}. In addition, the $^{207}$Pb/$^{206}$Pb age of a troilite nodule in an 
IVA iron meteorite shows that these irons cooled within 2-3 Myr of the formation
of Calcium-Aluminum-rich Inclusions (CAIs)
\citep{Blichert-Toft10}. A differentiated body that was small enough to have 
solidified and cooled in 2-3 Myr could not have retained sufficient heat released 
by $^{26}$Al to have melted. Thus the antiquity of the IVA iron meteorite requires an 
early collision to separate core and mantle material. The evidence that the IVA 
parent body was disrupted within 2-3 Myr of CAI formation, when chondrites parent 
bodies were accreting, supports the idea that the iron meteorites are derived from 
bodies that did not form in the asteroid belt.

Another line of support for the model by \citet{Bottke06} comes from the fact that
iron meteorites have concentrations of moderately volatile siderophile elements 
that are lower than those in chondrites. For example, the most depleted groups, 
IVA and IV, have Ge/Ni and Ga/Ni ratios that are $10^{-2}$ to $10^{-4}$ times those in chondrites. 
These depletions are probably due to early accretion in a hot solar nebula, as the 
alternative explanation, the impact volatilization during mantle removal, fails to 
explain the large depletions in group IVB, which crystallized before mantle loss 
\citep{Yang10a}. Thermal models for the solar nebula and calculated condensation 
temperatures for Ga, Ge, and the other volatile siderophile elements suggest that 
the parent bodies of the IVA and IVB irons may have accreted 0.3 Myr after CAI formation 
at a distance of 0.9 AU from the Sun \citep{Bland10}. Group IIAB and IIIAB irons 
have higher concentrations of Ga and Ge suggesting later accretion (0.5-1.0 Myr) 
after CAIs at $1.0\pm0.1$ AU. So, the chemical compositions of iron meteorites provide 
strong support for their origin in bodies that accreted closer to the Sun and earlier 
than the chondrites parent bodies. 

Given these arguments favoring derivation of igneously differentiated asteroids and
meteorites from the terrestrial planet region, it is important to test the \citet{Bottke06}
model in more detail as it neglected many processes that would have affected the growth
of planetesimals and scattering of their fragments into the asteroid belt. Here we focus
on the effects of the giant planets, which were completely ignored by \citet{Bottke06}.
By not including these effects, this model implies that iron meteorites parent bodies have 
to be melted to form cores, solidified, 
broken open, and emplaced in the asteroid belt before the growing, or fully grown giant planets 
destabilize the asteroid belt, ending accretion there and remove protoplanetary objects.
However, it is widely accepted that during the time that in the inner solar system, runaway 
growth among planetesimals was in progress to form planetary embryos
\citep{Safronov69,Greenberg78,Wetherill89,Weidenschilling97,Kokubo02}, 
giant planets were growing in the outer parts. As a result,
the dynamics of planetary embryos, which are the facilitators of the 
out-scattering and capture of planetesimals, could have been affected by these objects.
This has been reported in the works of many authors including 
\citet{Wetherill96}, \citet{Petit01}, \citet{Chambers02}, \citet{Chambers03}, \citet{Levison03},
\citet[][2009]{Raymond04}, \citet[][2007]{OBrien06}, and \citet[][2010]{Morishima08} 
who have shown that the gravitational perturbation 
of a fully formed Jupiter profoundly affects planetesimal scattering in the asteroid belt.
A more complete model of the origin of the parent bodies of iron meteorites has to 
include the effects of growing giant planets as well. It would be important to examine 
whether as the giant planets grow, sufficient numbers of the parent bodies of iron 
meteorites will in fact be injected into the asteroid belt. This paper presents the results 
of the first attempt in modeling this process.

Our goal is to show the proof of concept. That is, we would like to show that 
a planet in an outer orbit will in fact affect the rate of the scattering
of differentiated planetesimals into the inner asteroid belt.
At the time that the model by Bottke et al was proposed, it was considered that
the timescale for the formation of giant planets would be as long as in the original 
core-accretion scenario \citep[i.e., $\sim$10 Myr,][]{Pollack96}. As such, the iron meteorite parent 
bodies could form and be scattered into the asteroid belt before Jupiter would be large enough to deplete the 
mass of the entire belt. However, if the time of giant planet formation is short, as in the disk 
instability model \citep{Boss00,Boss01,Mayer02}, or comparable with the average lifetime of disks 
around young stars (e.g., $\sim$3 Myr)\footnote{The observational estimates of the lifetimes of 
disks around young stars suggest a lifetime of 0.1-10 Myr, with 3 Myr being the age for which half 
of stars show evidence of disks \citep{Strom93,Haisch01,Chen04}.} as in the models by
\citet{Rice03}, \citet{Alibert04}, \citet{Hubickyj05}, \citet{Lissauer09}, and \citet{Movshovitz10}, 
the perturbing effect of giant planets 
during their growth may not be negligible. The goal of this paper is show the latter concept.
As mentioned before, a realistic model requires the simulations to be carried out while giant 
planets are growing. This is a complicated task that requires complex computational programming
and long time simulations. However, before taking up on such a project, which is beyond the scope
of this paper, it proves useful to determine if the effect is large enough to be worthy
of the effort. It would also be helpful to determine a minimum mass for the giant planet for 
which its perturbation will be non-negligible. The latter will allow us to estimate a timescale
for the formation of this planet that would be consistent with the time of the formation of the 
parent bodies of iron meteorites. 
For these reasons, in this paper, we consider the planet to have been fully formed, and
carry out simulations for different values of its mass. Also, in order to be able to compare
our results with those of \citet{Bottke06}, we run simulations for 10 Myr. 

The outline of this paper is as follows. In Section 2, we present our model and the 
initial set up of the simulations. A detail of our numerical integrations and results 
is given in Section 3. Section 4 concludes this study by discussing the implications 
of the results to the origin of the parent bodies of iron meteorites, and their connections 
to the time of giant planet formation.

\section{The Model and Initial Set Up}

Since we are interested in portraying the effect of an outer
planet on the out-scattering of planetesimals from the terrestrial planet region into the
inner asteroid belt, we considered a heuristic model consisting of 
the Sun, a disk of protoplanetary embryos, and a fully formed planet. The disk included over one 
hundred Moon- to Mars-sized objects and more than 1200 planetesimals. We
randomly distributed planetary embryos between 0.5 AU and 4 AU with mutual 
separations of 3-6 Hill radii (figure 1). The masses of these objects were chosen to 
increase with their semimajor axes ($a$) and the number of their mutual Hill radii ($\Delta$) 
as ${a^{3/4}}{\Delta^{3/2}}$. The total mass of the protoplanetary disk was 
approximately 4 Earth-masses, 
and its surface density, normalized to 8.2 g cm$^{-2}$ at 1 AU, was considered to be 
proportional to $r^{-3/2}$ \citep{Weidenschilling77,Hayashi81}. We assigned a randomly 
chosen eccentricity between 0 and 0.05 to each protoplanet, and assumed that the initial 
orbital inclinations of these objects were 0.1 degrees. 

Following \citet{Bottke06}, the planetesimals were considered to be massless 
particles. Since we are interested in the out-scattering of these objects from the 
region interior to 2 AU, we only focused on that region and randomly distributed 
planetesimals between 0.5 AU and 2 AU. Similar to the embryos, the eccentricities of 
these bodies were chosen randomly from the range of 0 to 0.05, and their initial 
inclinations were taken to be 0.1 degrees. 

The planet was considered to be in the current orbit of Jupiter. As explained before,
we did not model the actual growth of the planet. Instead, we assumed that
the planet was fully formed and carried out simulation for different values of its
mass ranging from 0.1 to 300 Earth-masses.

Prior to describing the results, it is important to note that because in our simulations, 
planetesimals were treated as test particles, the mutual interactions between these 
objects were ignored. The embryos, on the other hand, were in full gravitational 
interaction with one another. These objects were allowed to collide and merge, and 
they could also accrete planetesimals (the latter would not change their masses). 
Since the focus of our study is on identifying the effect of the outer planet on 
the out-scattering of planetesimals, to avoid computational complexities, we assumed 
that collisions between embryos were perfectly inelastic and resulted in the accretion 
of the colliding bodies. During the simulations, the perturbing effects of growing 
embryos were taken into account. However, in order to better portray the out-scattering 
of planetesimals, the changes in embryos' radii are not shown in the figures
(the symbols for these objects are all identical in size). 

We would also like to emphasize that in a realistic model, in addition to the
effects of the giant planets, other processes such as gas drag, dynamical friction,
breakage and re-accretion of colliding objects, and the effects of the size and spatial 
distribution of planetesimals and planetary embryos must also be taken into account.
However, such inclusive and detailed modeling of the evolution of asteroid belt is 
beyond the scope of our paper. Our goal is merely to demonstrate that the perturbation of 
a giant planet is an important effect which cannot be ignored. In that respect, as we 
explained above, our model is heuristic, and the results of our simulations are only 
comparable with similar models with no giant planets.

\section{Results of Numerical Simulations}

We numerically integrated the orbits of the planetesimals, protoplanets, and the 
outer planetary body using the hybrid integrator of the N-body integrations package 
Mercury \citep{Chambers99}. The timestep of integrations was set to 6 days, and following 
\citet{Bottke06}, simulations were carried out for 10 Myr. 

Figures 2 and 3 show the results of simulations for planetary masses of 0.5 and 
1 Earth-mass at 2, 5, and 10 Myr. (Graphs of the 
inclination vs semimajor axis show the same features as the eccentricity plots.) A
comparison between the state of the surviving planetesimals and planetary embryos  
for each planetary mass at similar times indicates that for small values of the mass of 
the planet, the perturbing effect of this object is practically negligible.  In these systems, 
the dynamics of planetesimals and their scattering to outer regions is primarily governed by 
their interactions with the planetary embryos. As shown in figure 2, the majority of the planetesimals
that were scattered into the inner asteroid belt were originally from the region of 1.5-2 AU
(29\% in the simulations of 0.5 Earth-mass and 28\% in the simulations of 1 Earth-mass).
This is an expected result that is also consistent with the results reported by \citet{Bottke06}. 
As explained later, planetesimals from the 1-1.5 AU region also contributed. However, the
contribution of these objects was small and appeared late during the simulations.

As expected, when the perturbing planet has a larger mass, its effect becomes more apparent.
Simulations have shown that systems with planetary mass of 10 to $\sim 50$ Earth-masses 
form a transitional regime in which the perturbing effect of the planet appears mainly
at late times during the integration. Figures 4 and 5 show a sample of the results. 
A comparison between the results shown in the top and middle panels of figure 4 with those of 
figure 2 indicates that although during the first 5 Myr, the dynamics of planetary embryos in the
two figures is slightly different, the overall dynamics of their planetesimals is closely similar.
This is primarily due to the fact that the perturbing effect of the planet will require some time
to reach the planetesimals interior to 2 AU. As the planet interacts with planetary
embryos, it excites the orbits of these objects, in particular those in the 
region of 3-4 AU, causing their orbital eccentricities to rise to high values. For a given planetary
orbit, this orbital excitation is enhanced by the mass of the planet.
Simulations show that while at the end of the 10 Myr integration, in a system with a 1 Earth-mass planet,
the orbital eccentricities of embryos at the outer region of the disk did not exceed 0.16, 
in a system with a 50 Earth-mass planet, the planetesimals eccentricities rose up 
to $\sim 0.6$. As a result, more than 30\% of the embryos between 2 AU and 4 AU in this system 
were ejected (i.e., their aphelion distances become larger than 100 AU).
The interaction among planetary embryos causes their dynamical excitation to spread to 
those in closer orbits which in turn affects the dynamics of planetesimals in those regions.
Our simulations show that for systems with a planetary mass in the transitional regime, 
this process may take close to 10 Myr (the bottom panels of figure 4).

In systems where the planet was as massive as a gas giant, the results were quite
dramatic. Figures 6 and 7 show the results for planetary masses of 100 and 300
Earth-masses. As shown in figure 6, planetary embryos at 3-4 AU are strongly disturbed
by the perturbation of the planet. During the first 2 Myr of integration, 50\% of these 
embryos were scattered out of the system. The rapid progression of dynamical excitation
to embryos in closer distances during this time caused 40\% of all embryos to leave the system. 
This number rose up to 58\% after 5 Myr and 70\% after 10 Myr. At this time,
almost all embryos in the region of 3-4 AU had been ejected, and only 38\%
of the original embryos in the region between 2 AU and 3 AU survived. 

The dynamical excitation and ejection of embryos had profound effects on the dynamics
and out-scattering of planetesimals. While similar to previous cases, the majority of the
planetesimals that were scattered to the inner asteroid belt, originated in the region of 
1.5-2 AU, the number of these planetesimals was drastically reduced. Unlike the simulations
of figure 2 where $\sim 28\%$ of the planetesimals from the region 1.5-2 AU reached
the inner asteroid belt (this number was 21\% for the transitional case, figure 4), 
in the simulations with a giant planet, this number dropped to 12\%. 
This decreasing trend is also 
observed in the out-scattered planetesimals from the 1-1.5 AU region. Unlike the simulations
of figures 2 and 4 where in average close to 5\% of these objects were scattered into the region outside
2.1 AU, the highly excited embryos in the system with a giant planet caused only 
slightly over 1\% of the planetesimals from the region 1-1.5 AU to reach the inner asteroid 
belt. In none of our simulations, planetesimal from 0.5-1 AU reached distances larger 
than 1.6 AU.

\section{Implications of the Results}

As mentioned earlier, a very important implication of the model by \citet{Bottke06} 
is that in this model the destruction of the parent bodies of iron meteorites 
occurs in the terrestrial planet region and not in the asteroid belt. 
The small bodies that are generated in this way will then be scattered into the inner
asteroid belt through their interactions with planetary embryos.
As shown by our simulations, depending on its mass,
an outer planet may play an important role in this process. 
The perturbation of this object may directly affect the dynamics of planetary embryos which
will subsequently affect the interactions of these objects with planetesimals and  
the efficiency of their back-scattering from the terrestrial planet region into the inner 
asteroid belt. Integrations 
point to 50 Earth-masses as the lower limit for which the perturbation of the outer planet 
can no longer be ignored. In systems with a smaller planetary mass, 
particularly those with planets smaller than 10 Earth-masses, the dynamics of planetary embryos
is primarily driven by their interactions with one another. In these systems, 
the efficiency of the scattering of planetesimals from terrestrial region into 
the inner asteroid belt increases as planetary embryos collide and grow to larger sizes. 
As the latter is a slow process, the implantation of planetesimals in the inner 
asteroid belt in these systems may require a long time. 

It is important to note that
the time of the delivery of differentiated planetesimals to the inner asteroid belt
has to be consistent with cosmochemical data.
This time is  subject to two limiting factors; 1) the 
time of the cooling of iron meteorites, and 2) the time that it takes for the outer planet to 
grow to the critical mass of 50 Earth-masses.
At the time of the publication of the paper by \citet{Bottke06}, experimental data 
seemed to indicate that iron meteorites cooled over 100 Myr and therefore their parent 
bodies survived for that period of time \citep{Chabot06}. However, in the past few 
years, new cooling rate data and radiometric ages have shown that the destruction of 
iron meteorites parent bodies must have begun much earlier, perhaps as early as a few 
million years after the formation of CAIs \citep{Blichert-Toft10,Yang10b,Scott11}.
This implies that 
the process of the delivery of planetesimals to the inner asteroid belt
must occur during the first few million years after the formation of CAIs, and before the 
outer planet reaches 50 Earth-masses. To constrain this time, we note that
a comparison between the results of our simulations for all planetary masses show that
no significant differences exist between the dynamical state of planetesimals during the
first 2 Myr (top panels of figures 2, 4, 6). In all our simulations, 
the gravitational effect of the perturbing planet becomes readily apparent after 5 Myr, and is 
far more pronounced after 10 Myrs.
In other words, our simulations suggest that the period during which a significant 
number of planetesimals can be transferred from the terrestrial region into the inner
asteroid belt has to be within the first 5 Myr. This is a time that is also consistent with the time
of the destruction of the slow-cooling iron meteorites. 
Considering $\sim 50$ Earth-masses as the critical mass above which the perturbing effect
of the planet cannot be ignored, the statement above suggests that in a more realistic model,
where the effect of a growing planet in the orbit of Jupiter is considered, 
in order for the mechanism proposed by \citet{Bottke06} to efficiently deliver planetesimals 
from the region interior to 2 AU to the inner asteroid belt, the rate of the growth of the
giant planet has to be such that the planet does not reach 50 Earth-mass before the first 5 Myr.
In other words, a giant planet formation scenario in which a planet in the orbit of Jupiter 
%such as the disk instability model \citep{Boss00,Boss01,Mayer02}, 
%which a planet in the orbit of Jupiter may reach
may grow to 50 Earth-masses in less than $\sim 5$ Myr, may not be 
consistent with the time of the formation and cooling of the parent bodies of iron meteorites. 
It is important to caution that this result has been obtained using our simple model where
effects of the circularization of the orbits of embryos due to their tidal interactions with the
gaseous disk \citep{Tanaka04} and dynamical friction \citep{Adachi76} have been ignored. 
These effects, combined with the non-zero offset between the time-zero of the simulations
and the true age of the system, may slightly change the above-mentioned 5 Myr timescale.

Our simulations also show that, regardless of the mass of the outer planet, the largest
contribution to the scattered planetesimals from the terrestrial region into the inner asteroid 
belt ($<$ 2.5 AU), came from the 1.5-2 AU zone.
As mentioned in the previous section, the number of these objects 
decreases as the mass of the planet increases. This is primarily due to the fact that
in systems with larger planetary mass, planetary embryos (the facilitators of the
out-scattering of planetesimals) are dynamically excited and many of them are scattered out 
of the system. Simulations also indicate that in systems with 
small planets, where the effect of this body is negligible, a few planetesimals from the 
region of 1-1.5 AU reach the inner asteroid belt as well. Results show that 
the first of these objects 
entered the asteroid belt in approximately 5 Myr after the start of the simulation. Given
that the out-scattering of planetesimals is the result of their interactions with planetary
embryos, it is expected that in systems where 
the outer planet is massive and planetary embryos are ejected from the system at the early 
stage of its dynamical evolution, it may take even loner for planetesimals from the
region interior to 1.5 AU to reach the inner asteroid belt (see the bottom panels of figure 6). 
In other words, planetesimals from the region of 1-1.5 AU could have 
contributed to the parent bodies of iron meteorites. But this contribution is very small.
Any contribution from planetesimals in the region interior to 1 AU is, however, highly unlikely.

In closing, we would like to mention that the effect of the migration of the giant planet,
both during its growth and after its formation  
was not included in our simulations. While migration during formation, as suggested by 
\citet{Alibert04}, may not be consistent with the model discussed here (these authors
considered Jupiter to start its migration at 8 AU with a mass equal to a fraction of Earth
and to reside in its current orbit with its current mass after 1 Myr), post formation
migration may have a profound effect on the outcome of our simulations.
As suggested by \citet{Tsiganis05} and \citet{Walsh11}, giant planets in our solar system 
might have migrated after their formation \citep[also see][]{Morbidelli05,Gomes05}. The inward 
migration of Jupiter in these cases, increases the perturbing effect of this object on the 
dynamics of planetary embryos \citep{Gomes97}, and extends the range of its perturbation to 
deeper regions interior to 2 AU. It is expected that in this case, the efficiency of the 
back-scattering of planetesimals is reduced even more drastically as obtained from
our simulations.

\acknowledgments
We would like to thank the anonymous referee for critically reading our manuscript and for
his/her constructive suggestions. We acknowledge support from NASA 
Cosmochemistry program under the grant NNH08ZDA001N-COS. NH  also acknowledges support from NASA 
Astrobiology Institute (NAI) under Cooperative Agreement NNA04CC08A at the Institute for 
Astronomy (IfA), University of Hawaii (UH), and NASA EXOB grant NNX09AN05G.

\clearpage
\begin{figure}
\center
\vskip 20pt
\includegraphics[scale=0.5]{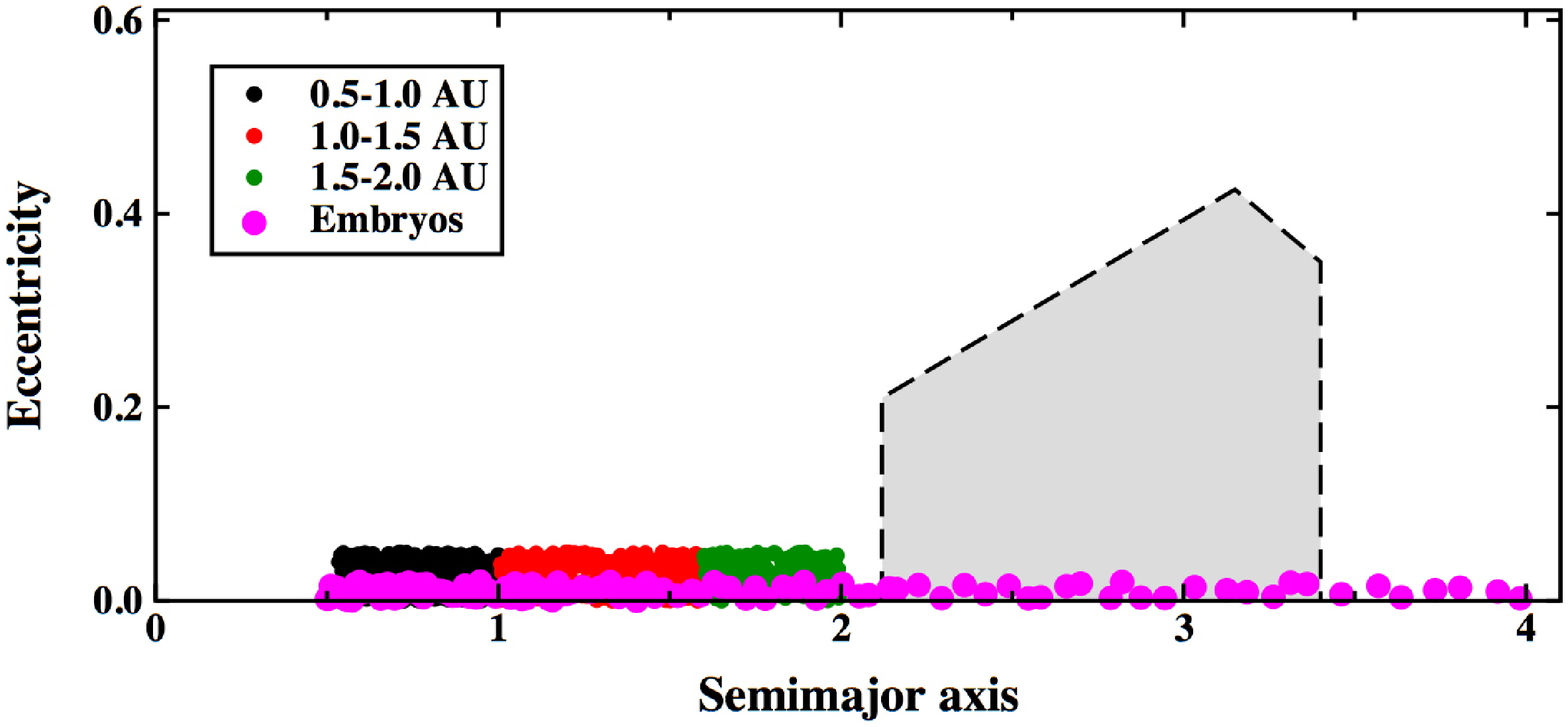}
\caption{Initial distribution of planetesimals (black, red, green) and planetary embryos
(pink). Note that because we are interested in the out-scattering of planetesimals from
the terrestrial region into the asteroid belt, we distributed these objects only between
0.5 AU and 2 AU. The gray area shows the asteroid belt.}
\end{figure}

\clearpage
\begin{figure}
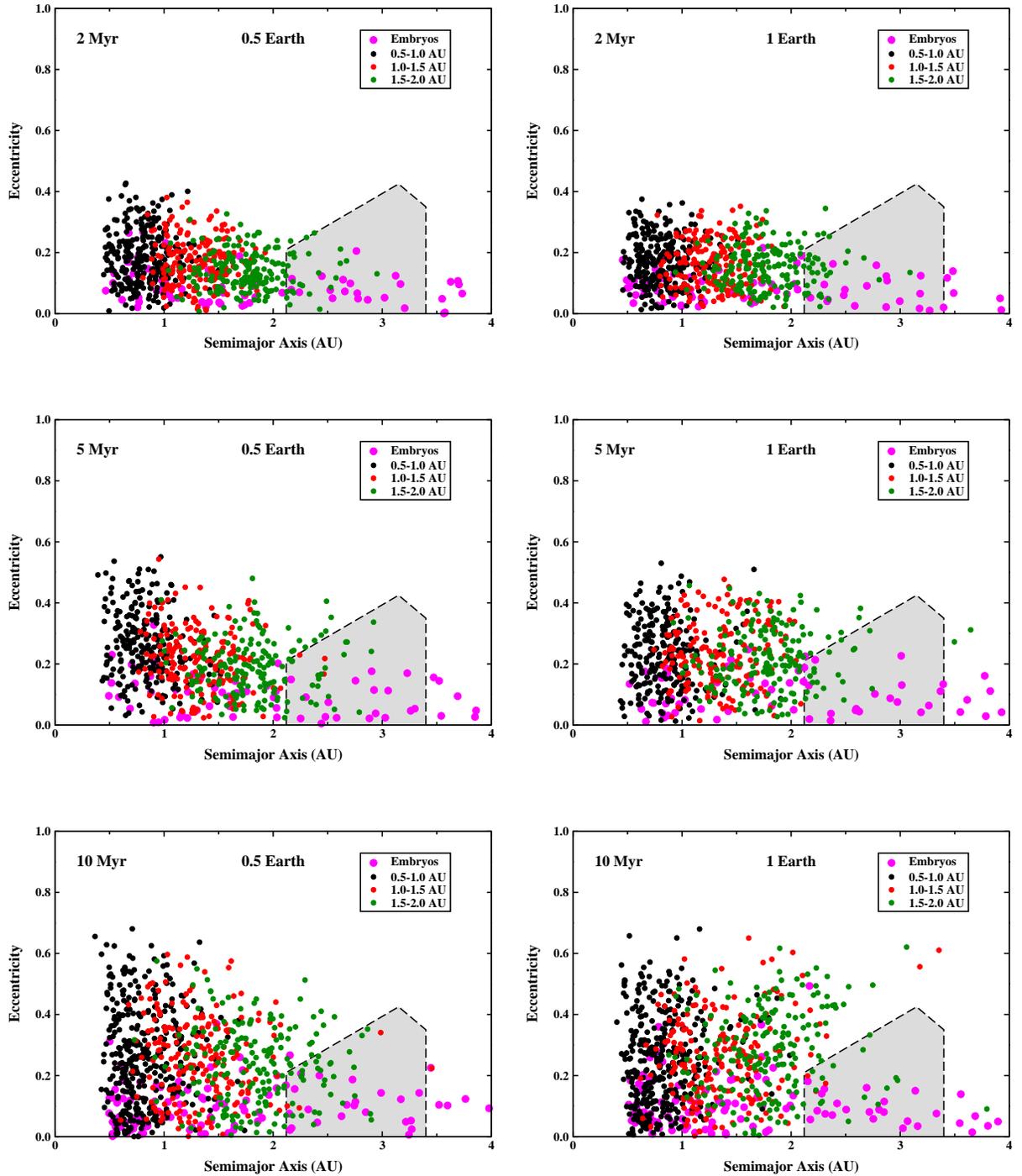

\center
\includegraphics[scale=0.32]{f2a.eps}
\hskip 10pt
\includegraphics[scale=0.32]{f2d.eps}
\vskip 20pt
\includegraphics[scale=0.32]{f2b.eps}
\hskip 10pt
\includegraphics[scale=0.32]{f2e.eps}
\vskip 20pt
\includegraphics[scale=0.32]{f2c.eps}
\hskip 10pt
\includegraphics[scale=0.32]{f2f.eps}
\caption{Snapshots of the dynamical state of planetesimals and planetary embryos
in a system with a 0.5 Earth-mass (left) and a 1 Earth-mass (right) planet
in the orbit of Jupiter.}
\end{figure}

\clearpage
\begin{figure}
\center
\includegraphics[scale=0.5]{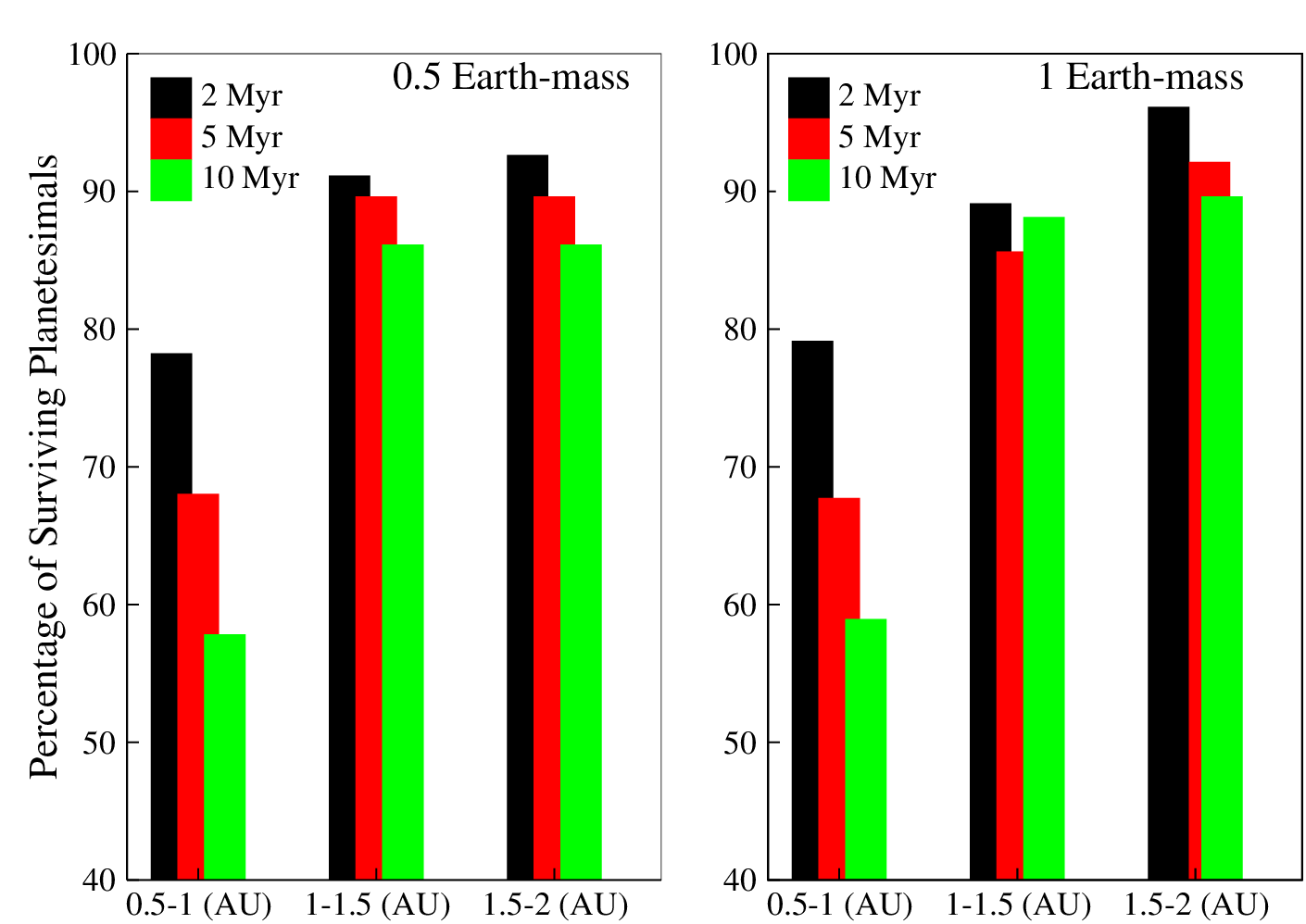}
\vskip 25pt
\caption{Percentage of survived planetesimals at the three zones 0.5-1, 1-1.5, and 1.5-2 (AU)
for a perturber with a mass of 0.5 and 1 Earth-mass.}
\end{figure}

\clearpage
\begin{figure}
\center
\includegraphics[scale=0.32]{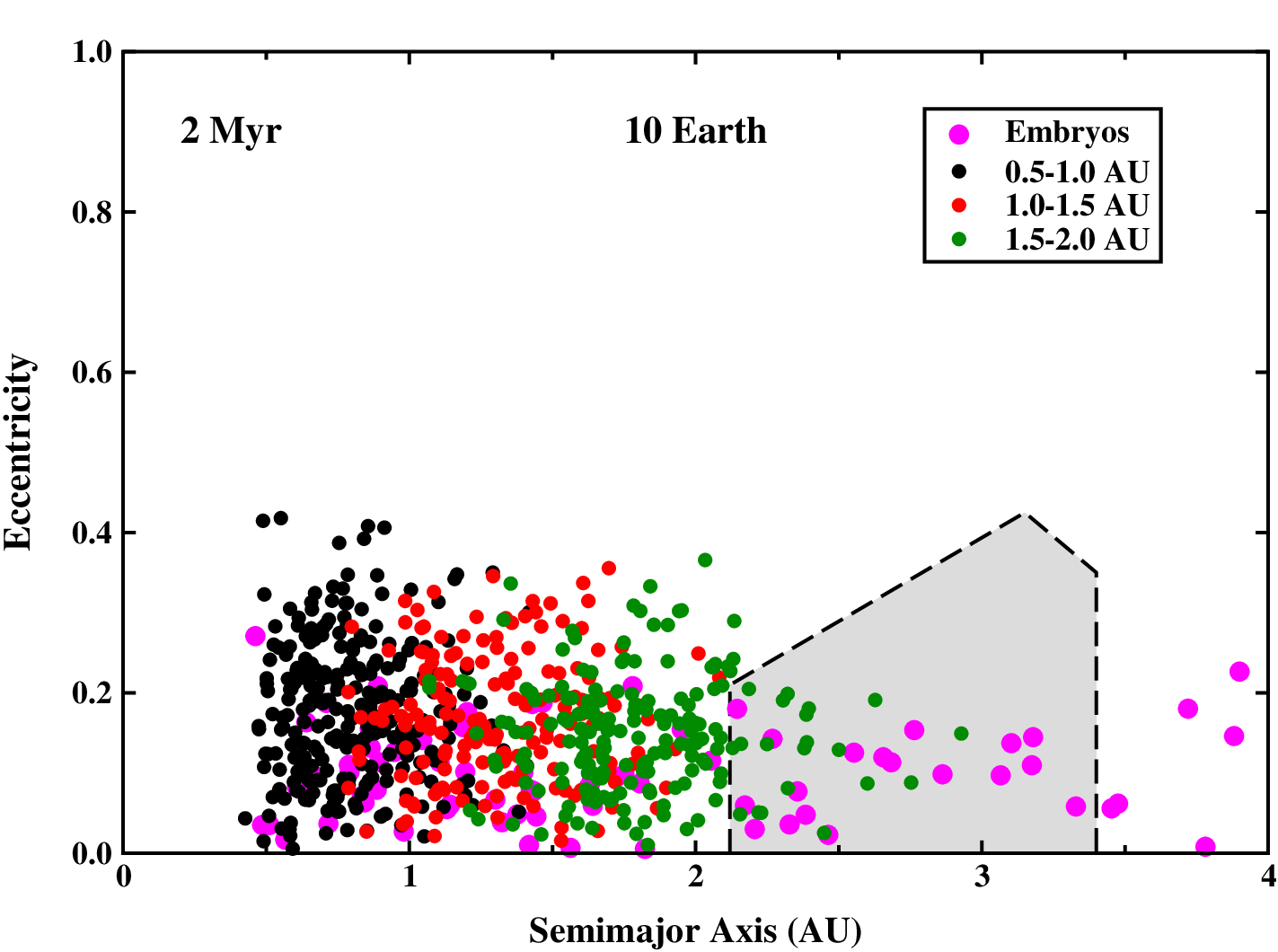}
\hskip 10pt
\includegraphics[scale=0.32]{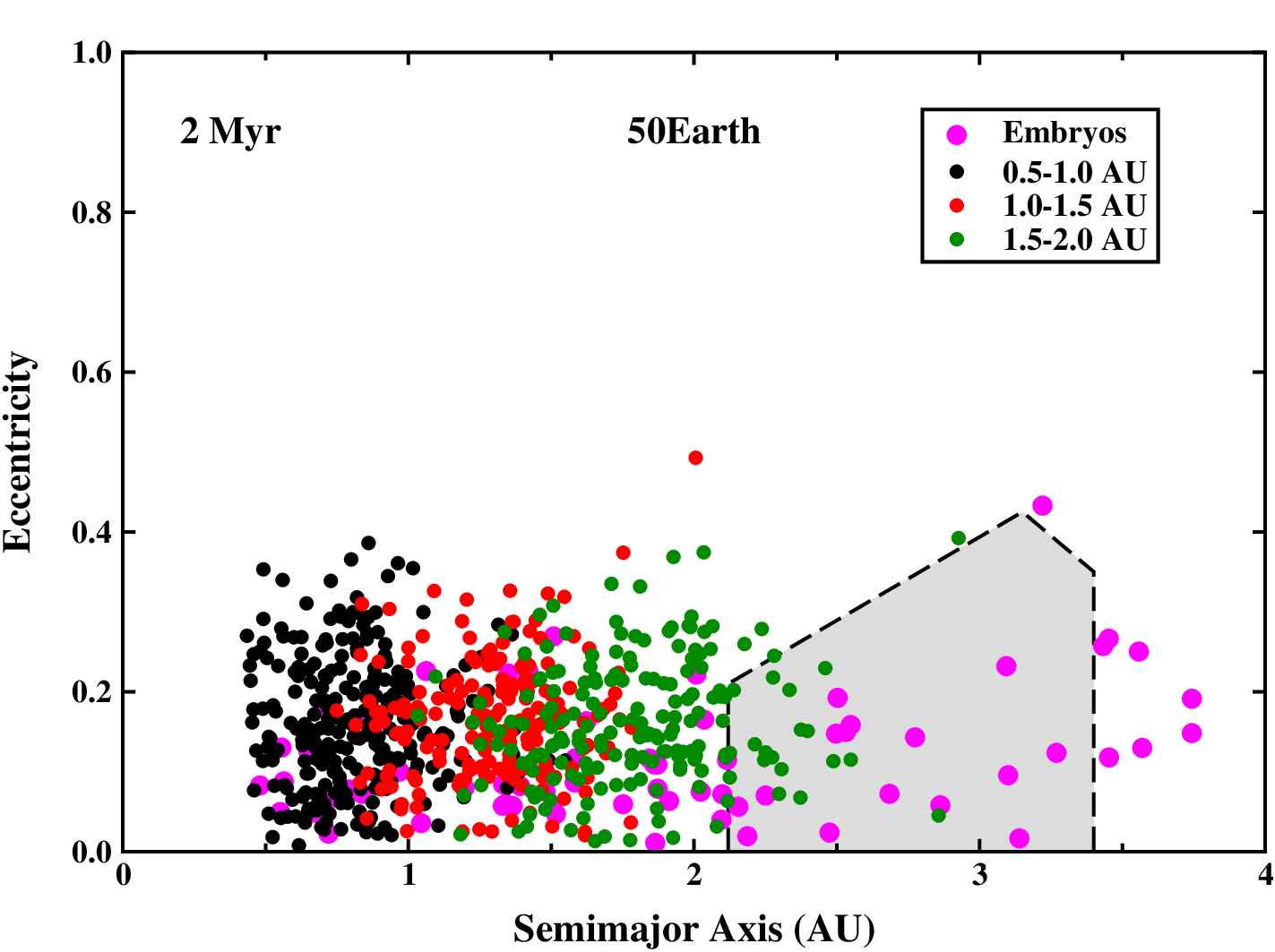}
\vskip 20pt
\includegraphics[scale=0.32]{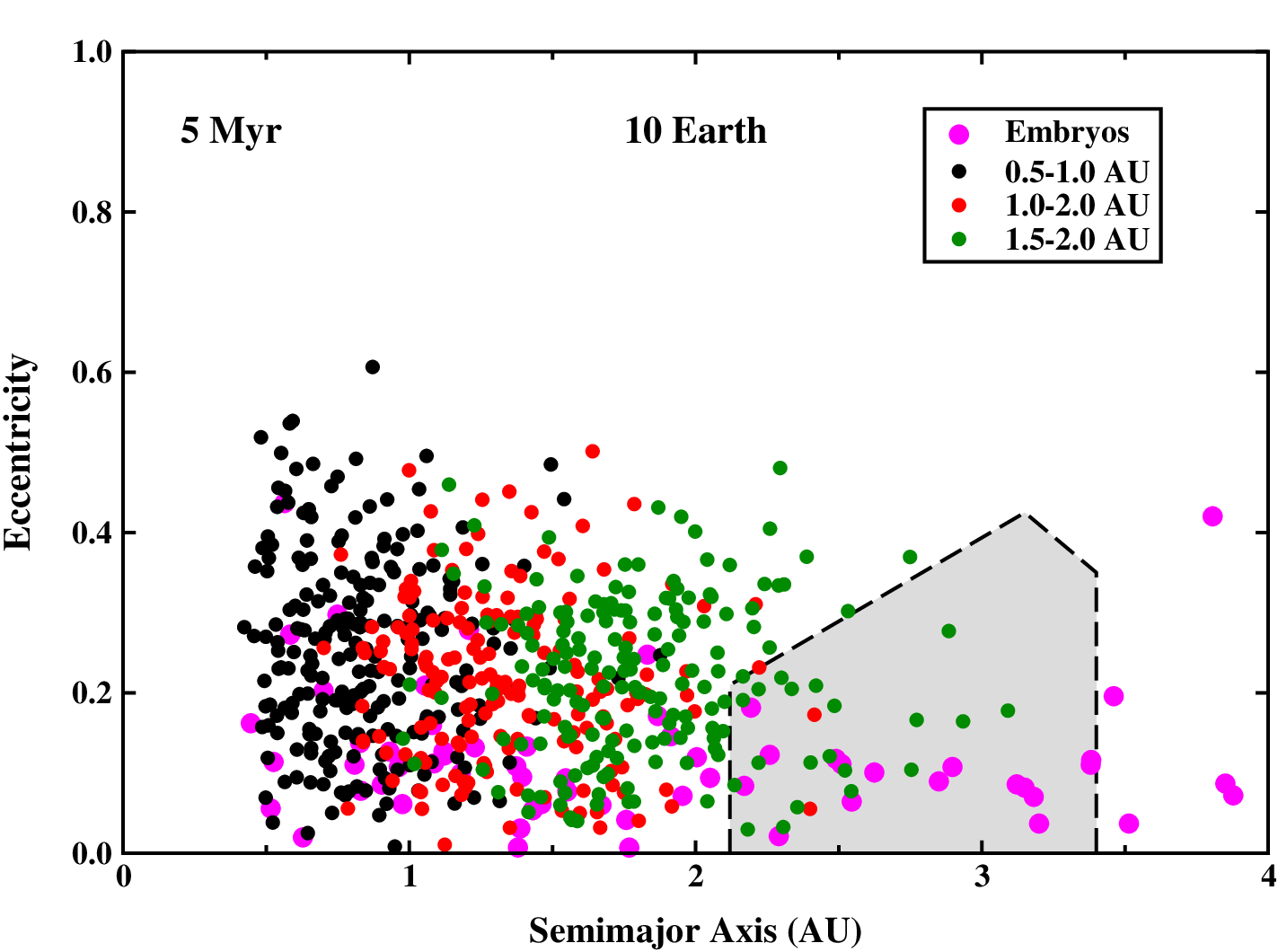}
\hskip 10pt
\includegraphics[scale=0.32]{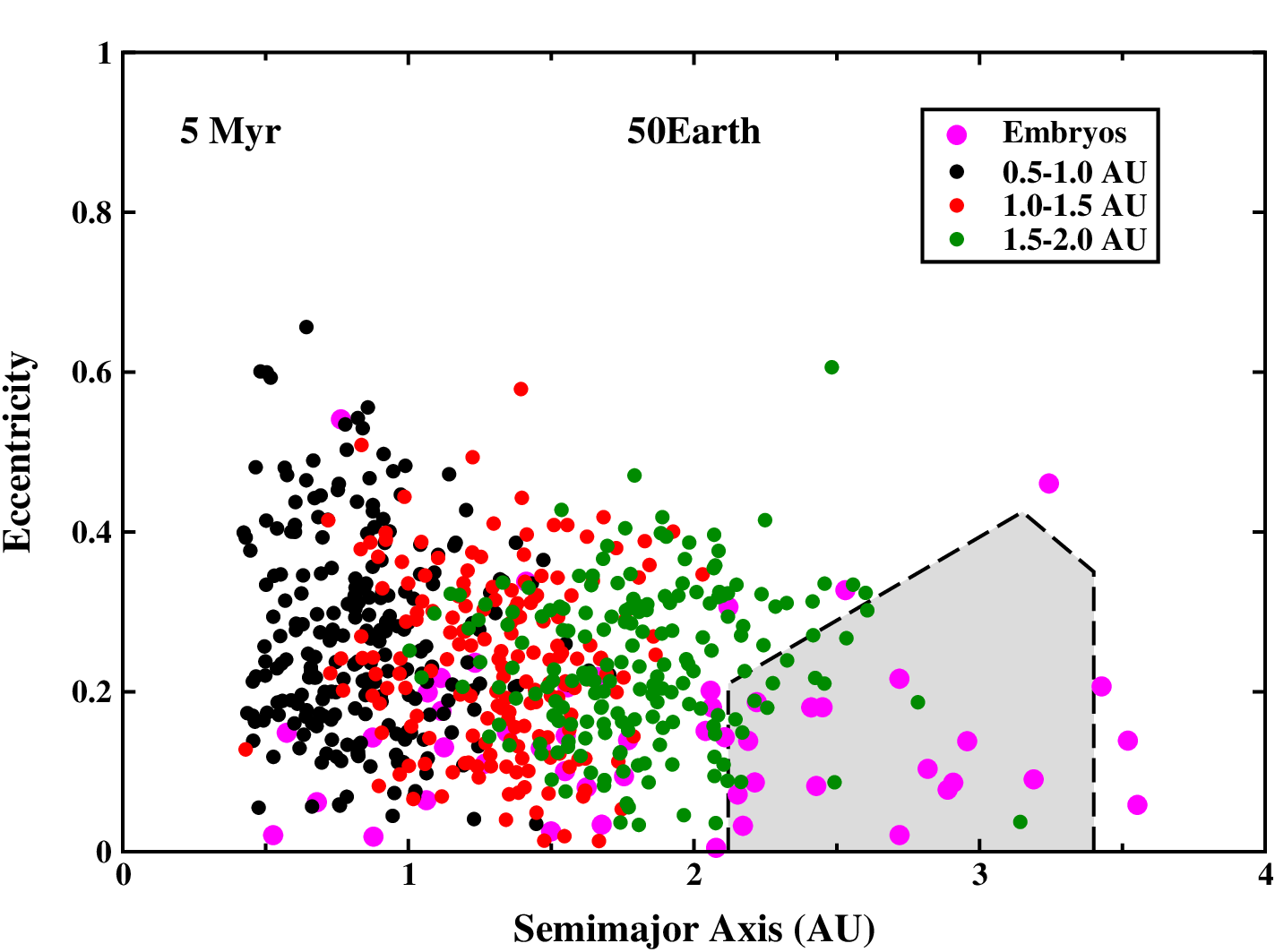}
\vskip 20pt
\includegraphics[scale=0.32]{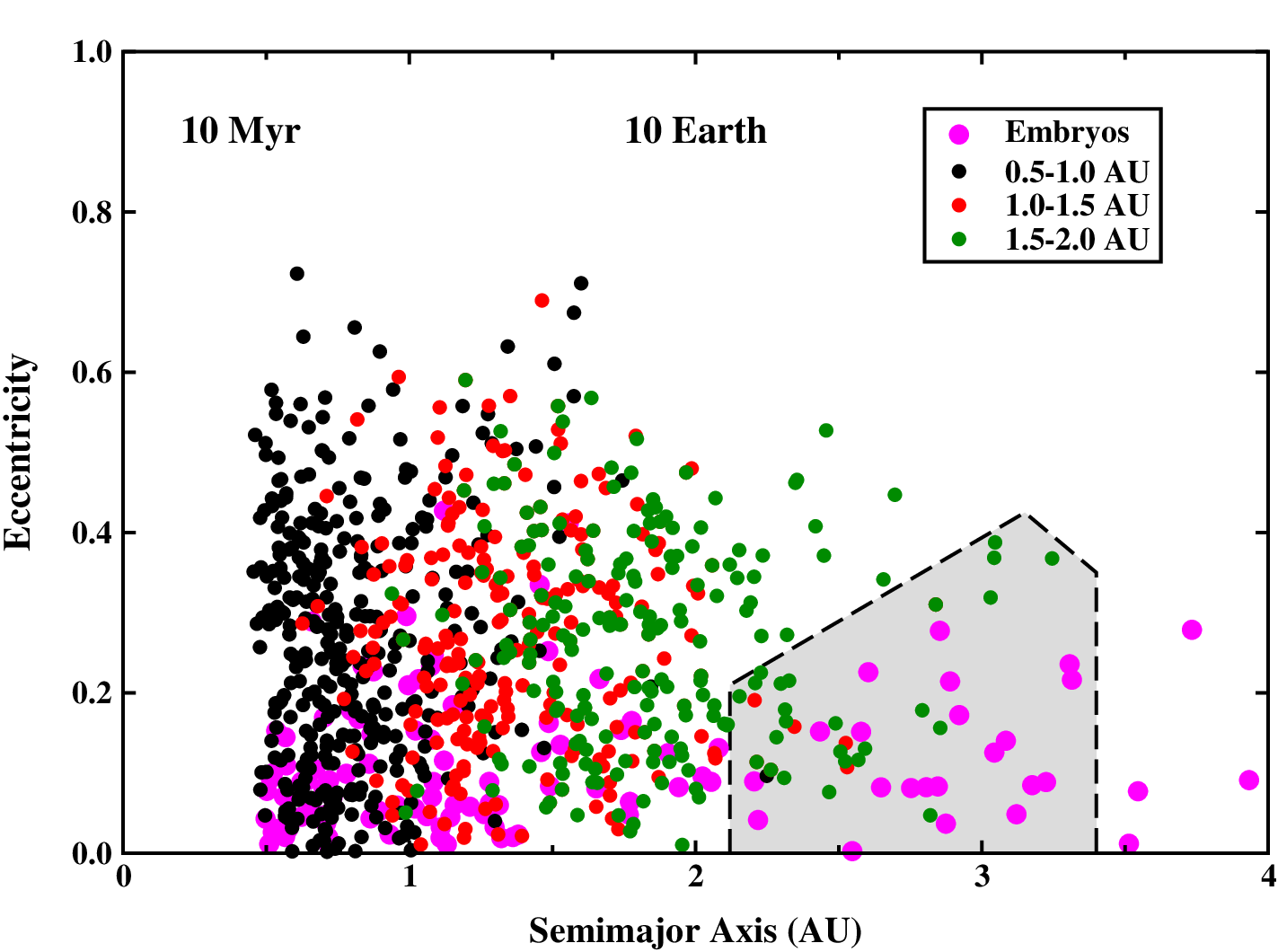}
\hskip 10pt
\includegraphics[scale=0.32]{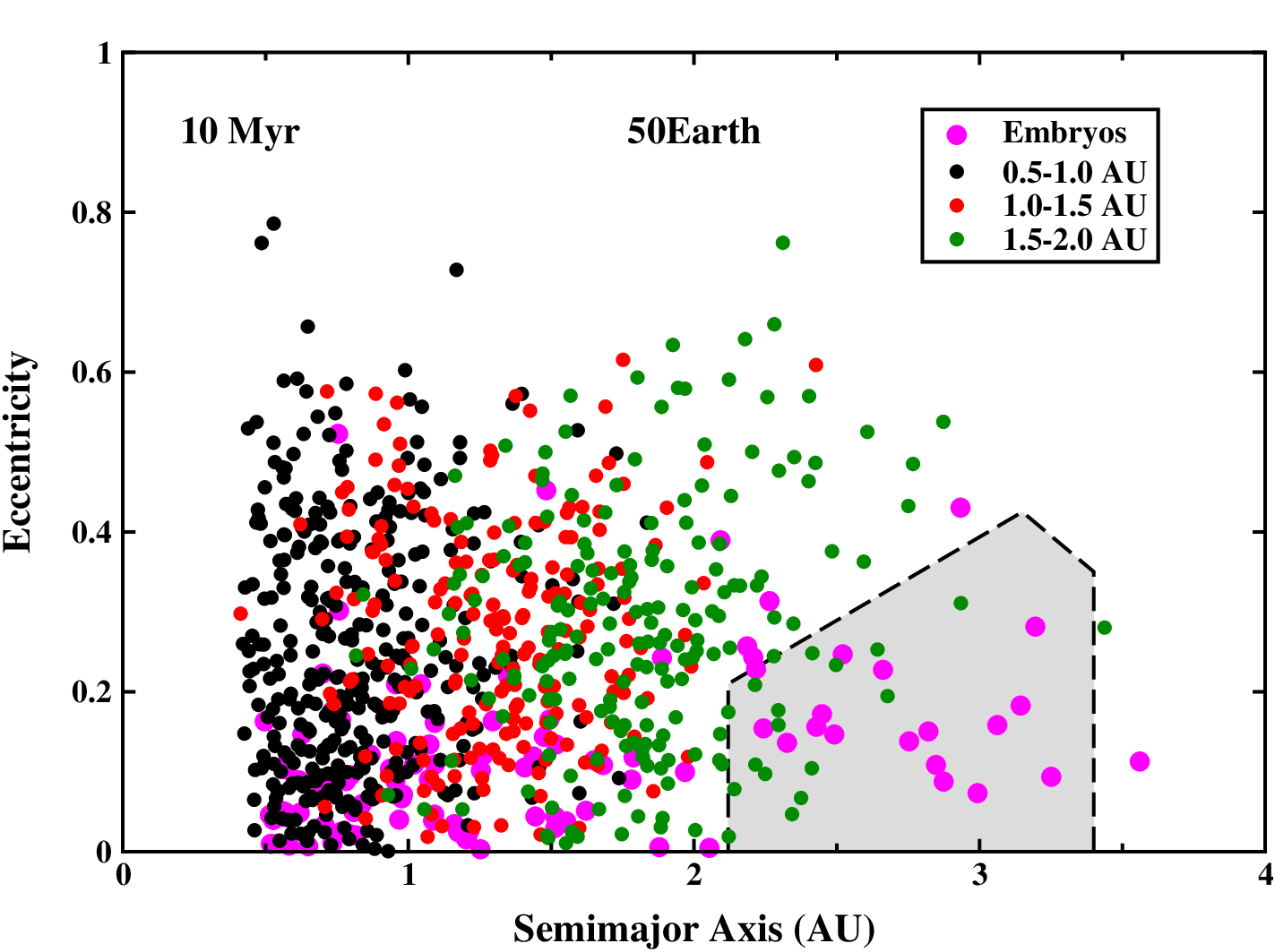}
\caption{Same as figure 2. The planet has a mass of 100 Earth-masses on the left
and 300 Earth-masses on the right.}
\end{figure}

\clearpage
\begin{figure}
\center
\includegraphics[scale=0.5]{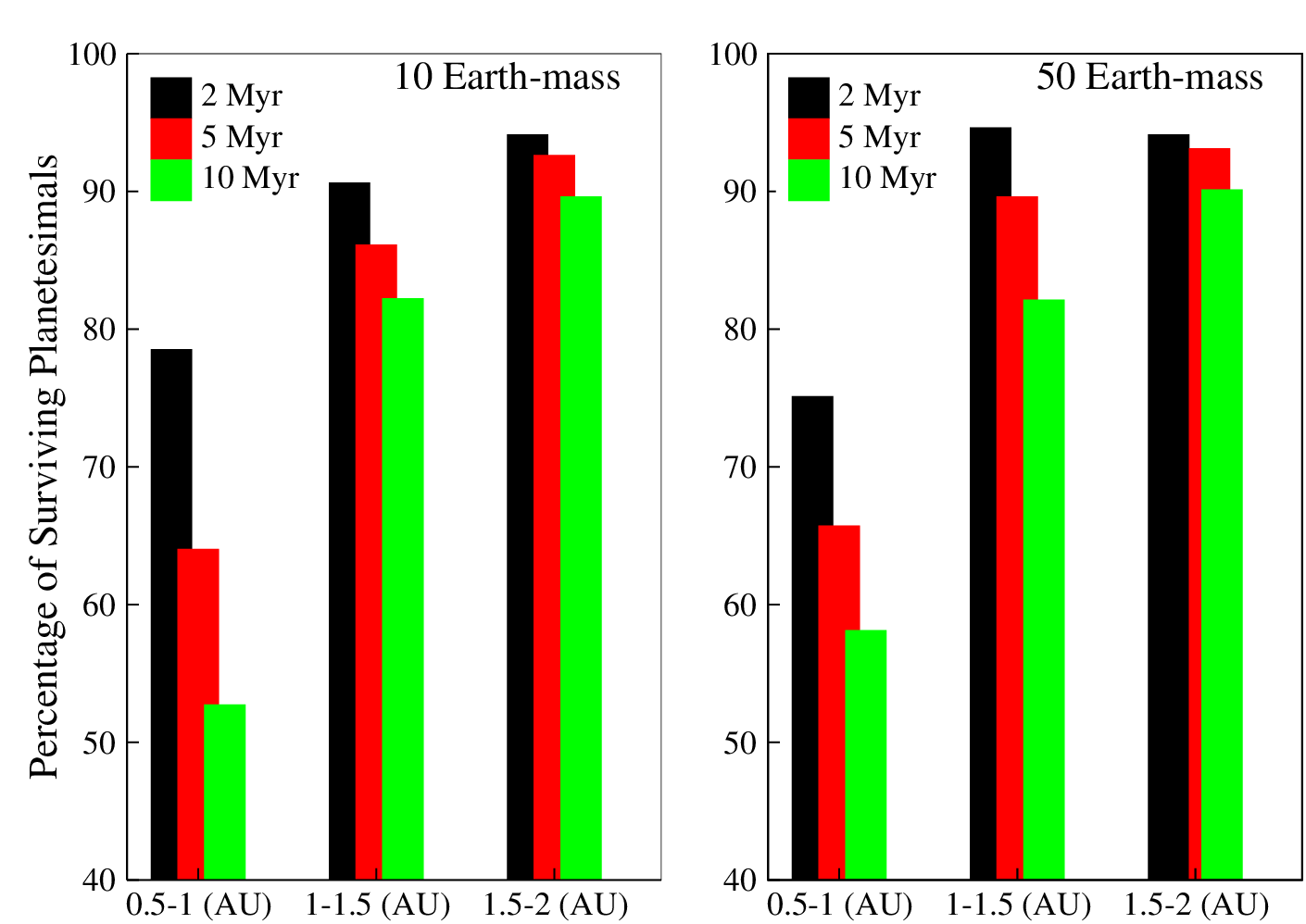}
\vskip 25pt
\caption{Percentage of survived planetesimals at the three zones 0.5-1, 1-1.5, and 1.5-2 (AU)
for a perturber with a mass of 10 and 50 Earth-masses.}
\end{figure}

\clearpage
\begin{figure}
\center
\includegraphics[scale=0.32]{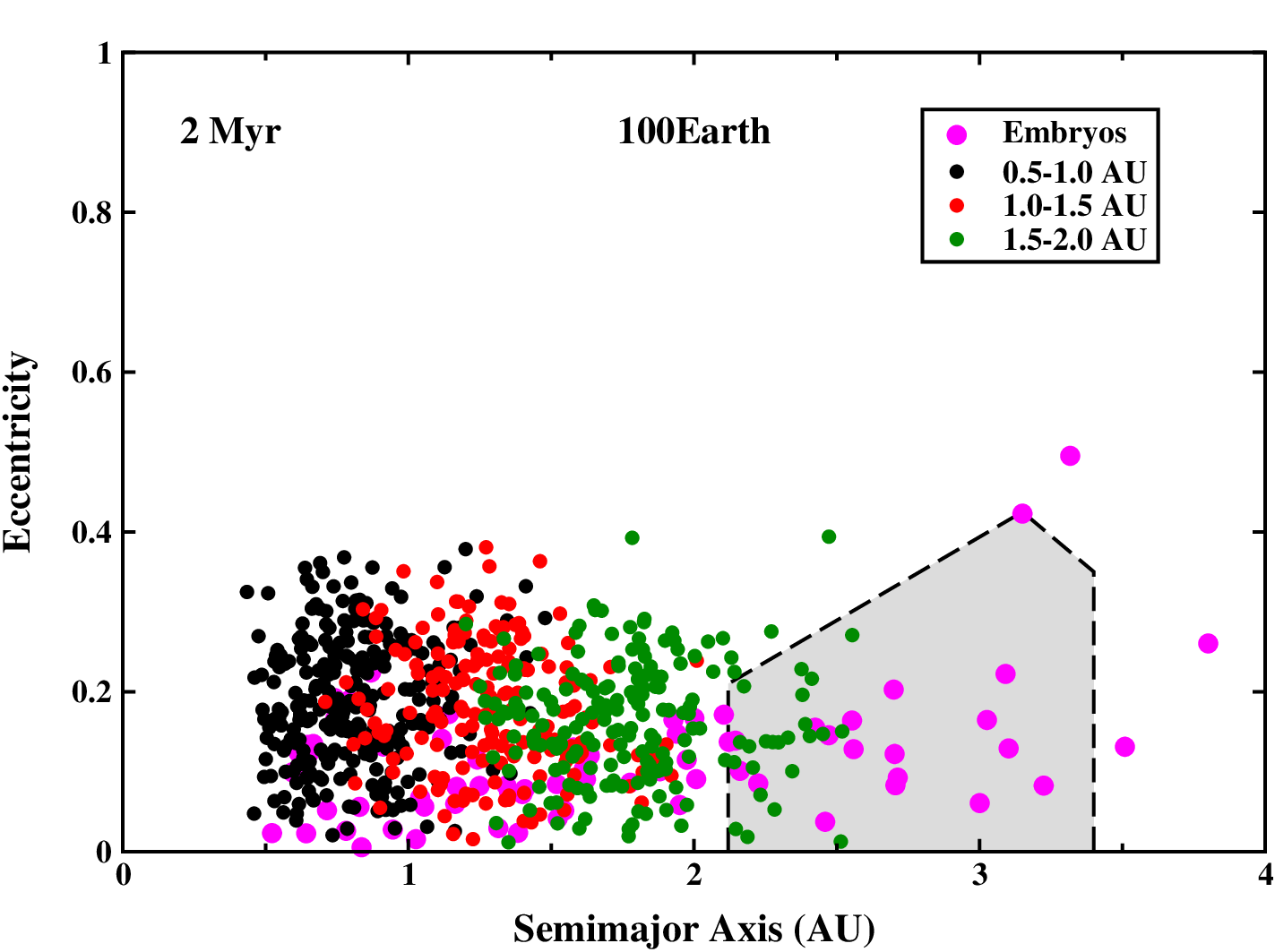}
\hskip 10pt
\includegraphics[scale=0.32]{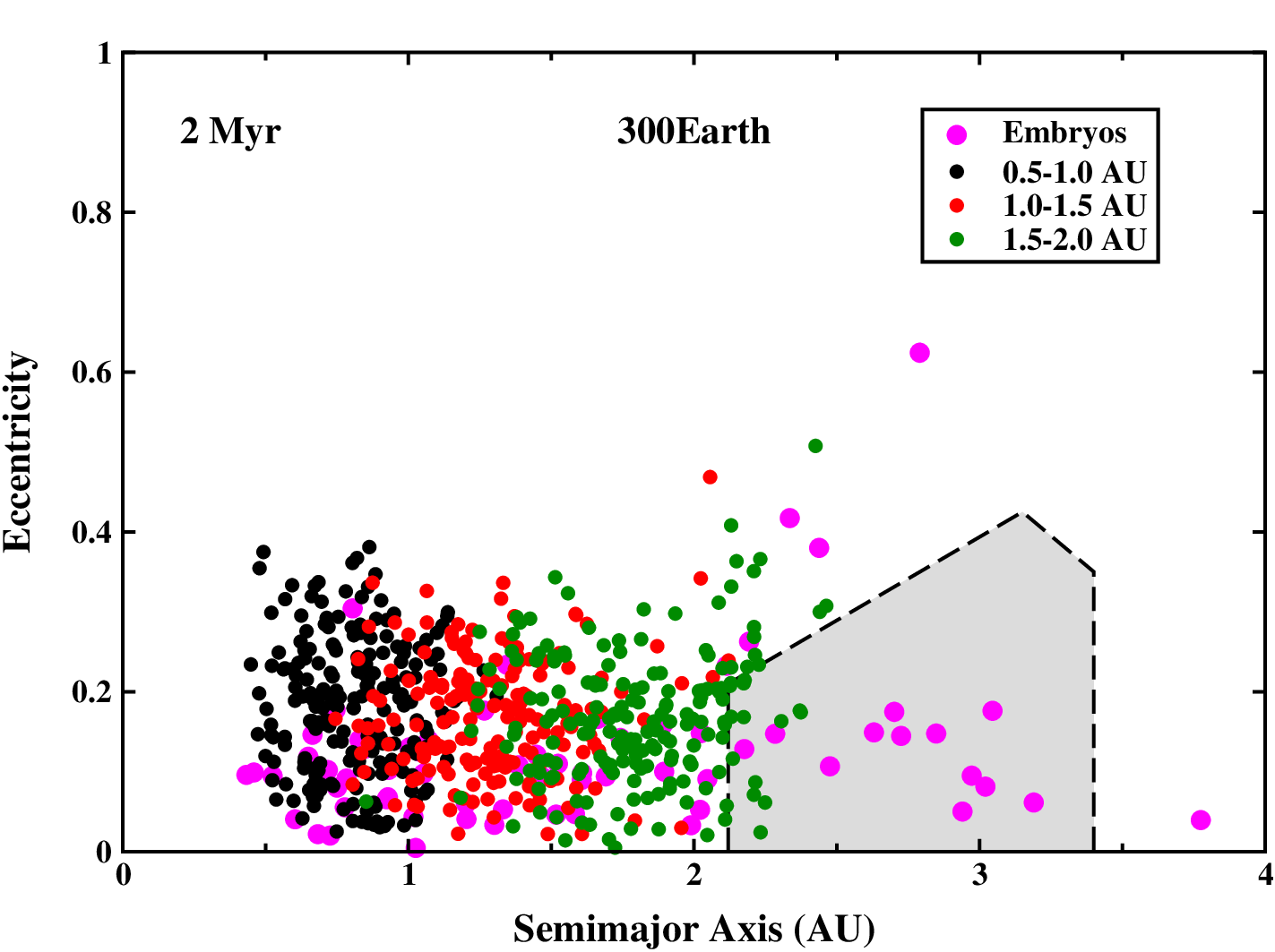}
\vskip 20pt
\includegraphics[scale=0.32]{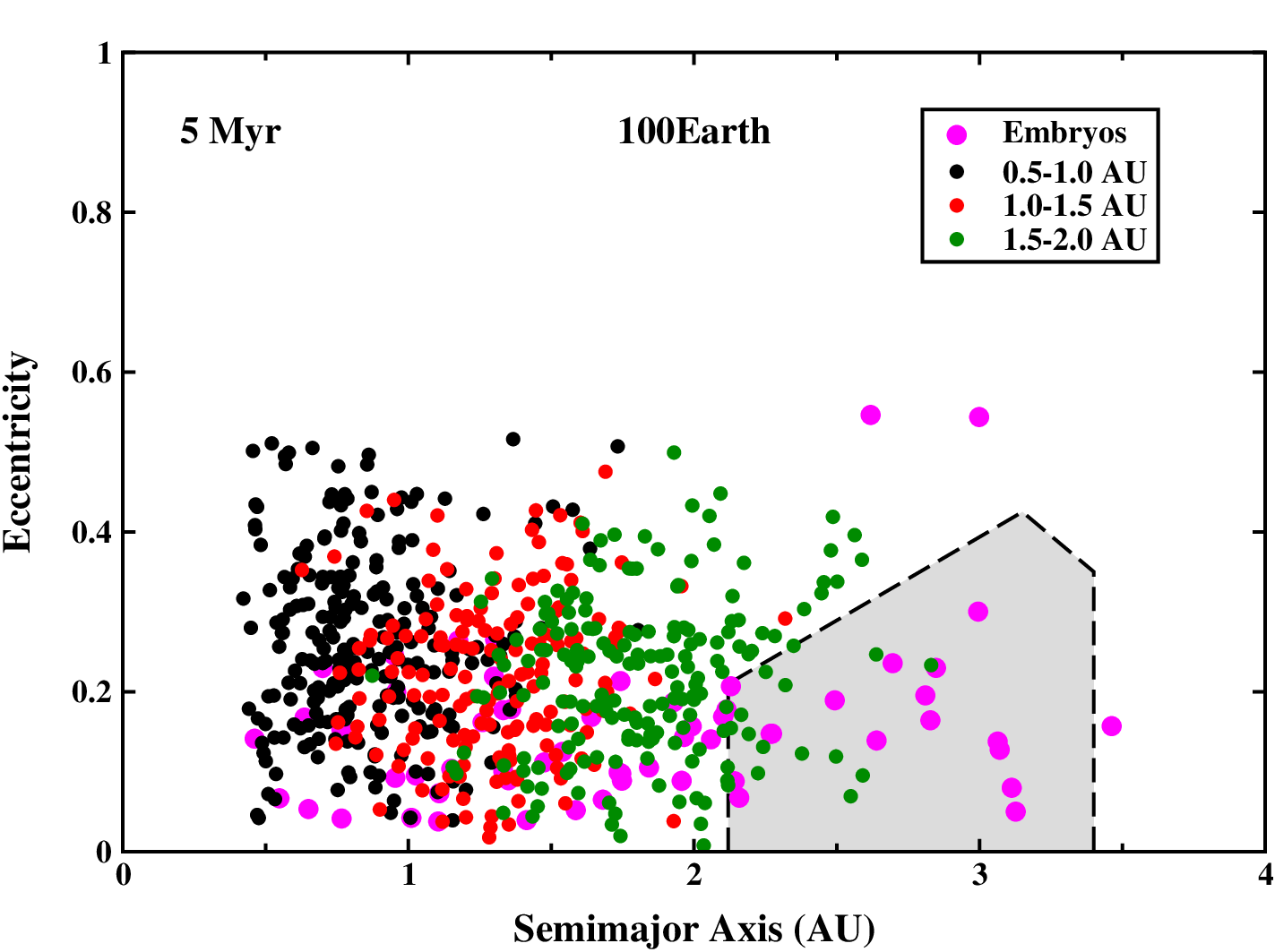}
\hskip 10pt
\includegraphics[scale=0.32]{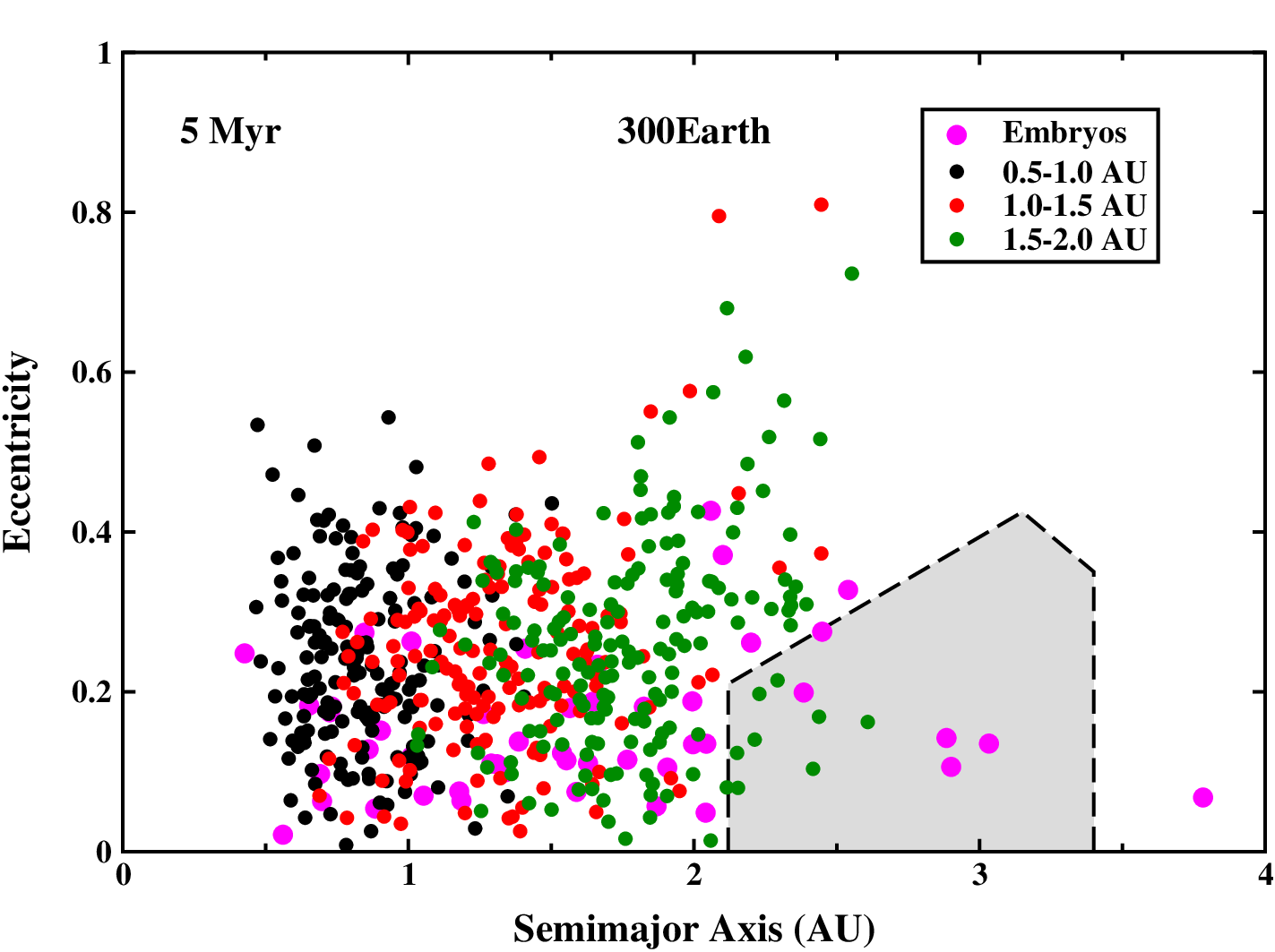}
\vskip 20pt
\includegraphics[scale=0.32]{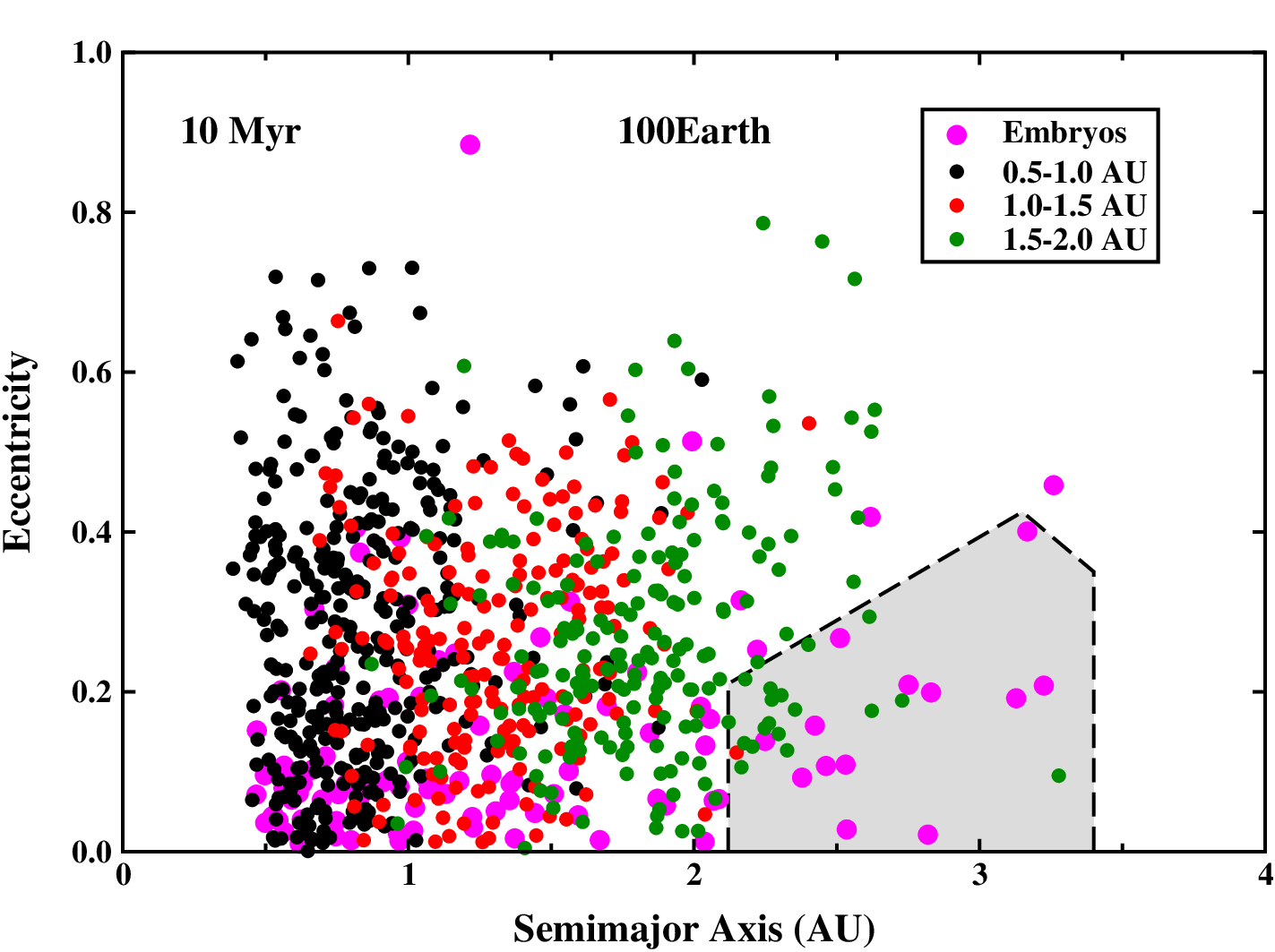}
\hskip 10pt
\includegraphics[scale=0.32]{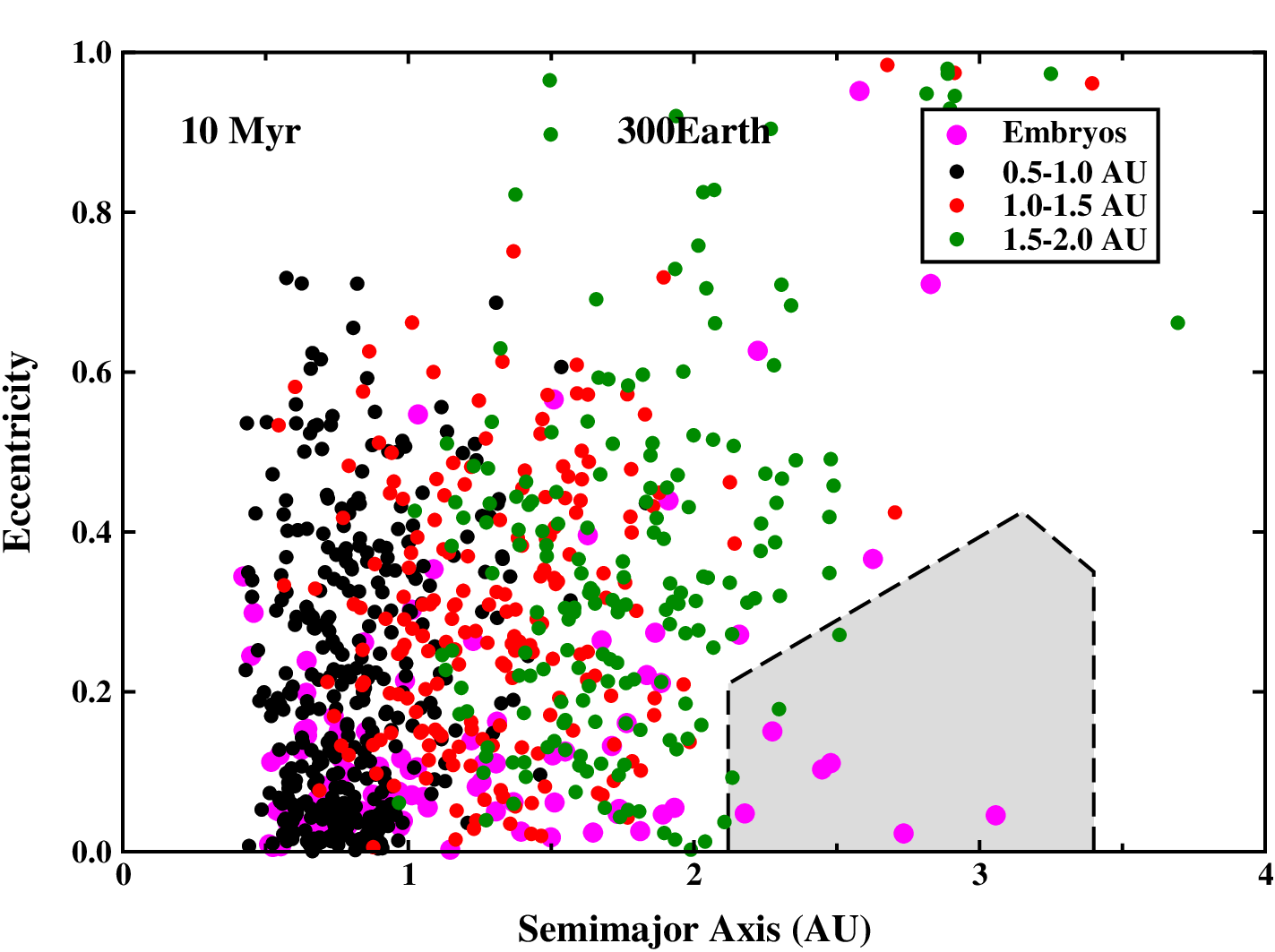}
\caption{Same as figure 2. The planet has a mass of 100 Earth-masses on the left
and 300 Earth-masses on the right.}
\end{figure}

\clearpage
\begin{figure}
\center
\includegraphics[scale=0.5]{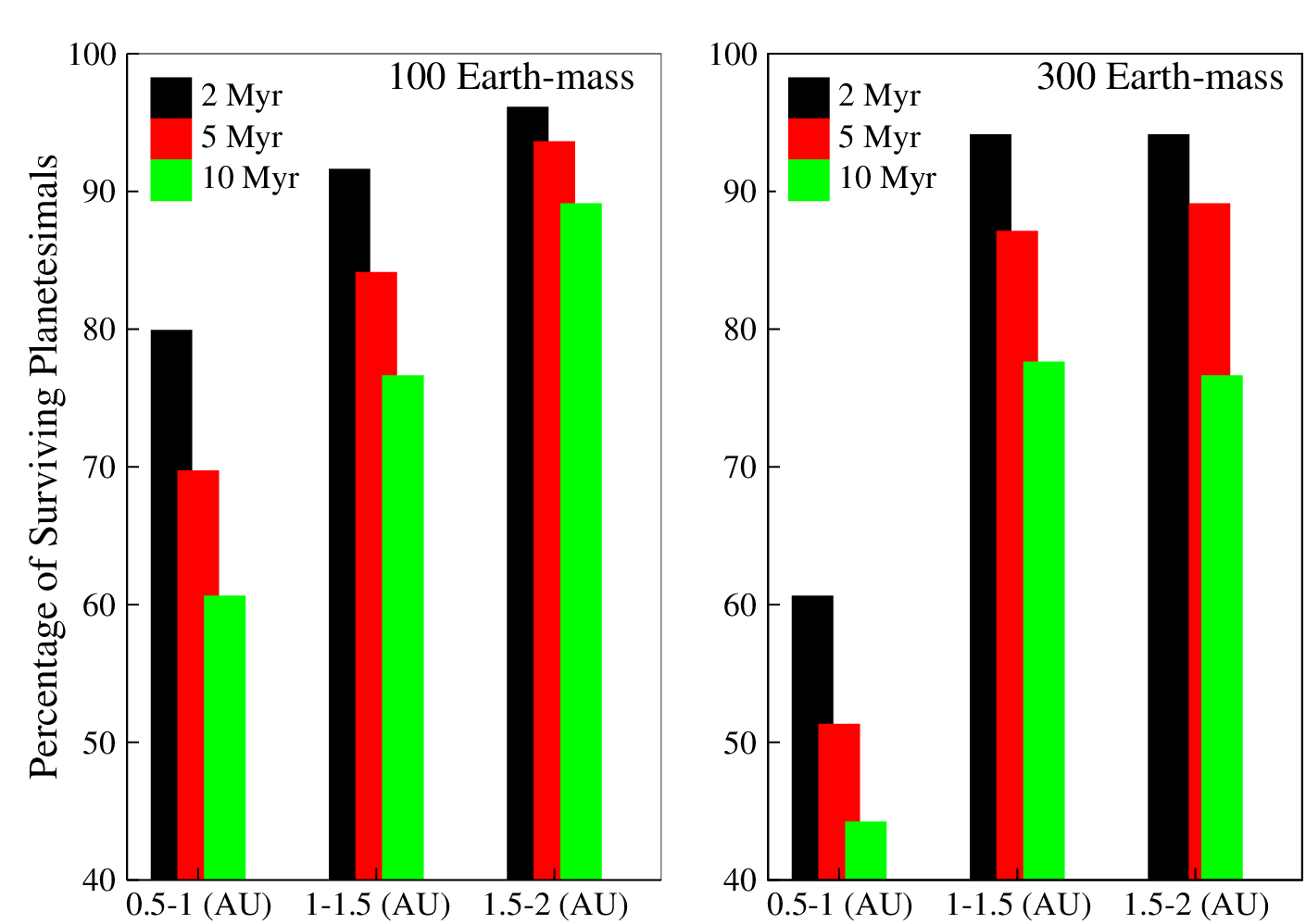}
\vskip 25pt
\caption{Percentage of survived planetesimals at the three zones 0.5-1, 1-1.5, and 1.5-2 (AU)
for a perturber with a mass of 100 and 300 Earth-masses.}
\end{figure}

\end{document}